\documentclass[apj,twocolumn]{emulateapj}

\submitted{Received 10 September 2008; Accepted by ApJ 21 December 2008}

\shorttitle{Major merger fraction at $z = 0.6$}
\shortauthors{L\'opez-Sanjuan et al.}

\begin{document}

\title{Robust determination of the major merger fraction at $z = 0.6$ in Groth Strip}
\author{Carlos L\'opez-Sanjuan\altaffilmark{1}, Marc Balcells\altaffilmark{1}, Cesar Enrique Garc\'{i}a-Dab\'o\altaffilmark{1,2}, Mercedes Prieto\altaffilmark{1,3}, David Crist\'obal-Hornillos\altaffilmark{1,4}, M. Carmen Eliche-Moral\altaffilmark{1,5}, David Abreu\altaffilmark{1}, Peter Erwin\altaffilmark{6}, and Rafael Guzm\'an\altaffilmark{7}}

\altaffiltext{1}{Instituto de Astrof\'{\i}sica de Canarias, Calle V\'{\i}a L\'actea s/n, E-38205 La Laguna, Tenerife, Spain}
\altaffiltext{2}{European South Observatory, Karl-Schwarzschild-Strasse 2, D-85748 Garching, Germany} 
\altaffiltext{3}{Departamento de Astrof\'{\i}sica, Universidad de La Laguna, E-38206 La Laguna, Tenerife, Spain}
\altaffiltext{4}{Instituto de Astrof\'{\i}sica de Andaluc\'{\i}a, Camino Bajo de Huétor, 50, E-18008 Granada, Spain}
\altaffiltext{5}{Departamento de Astrof\'{\i}sica y Ciencias de la Atm\'osfera, Facultad de C.C. F\'{\i}sicas, Universidad Complutense de Madrid, E-28040 Madrid, Spain}
\altaffiltext{6}{Max-Planck-Institut f\"ur extraterrestrische Physik, Giessenbachstrasse, D-85748 Garching, Germany}
\altaffiltext{7}{Department of astronomy, University of Florida, 211 Bryant Space Science Center, Gainsville, FL 32611-2055, USA}

\email{clsj@iac.es}

\begin{abstract}
We measure the fraction of galaxies undergoing disk-disk major mergers ($f_{\rm m}^{\rm mph}$) at intermediate redshifts ($0.35 \leq z < 0.85$) by studying the asymmetry index $A$ of galaxy images. Results are provided for $B$- and $K_{\rm s}$-band absolute magnitude selected samples from the Groth strip in the GOYA photometric survey. Three sources of systematic error are carefully addressed and quantified.  The effects of the large errors in  the photometric redshifts and asymmetry indices are corrected with maximum likelihood techniques.
Biases linked to the redshift degradation of the morphological information in the images are treated by measuring asymmetries on images artificially redshifted to a reference redshift of $z_{\rm d} = 0.75$. Morphological K-corrections are further constrained by staying within redshifts where the images sample redward of 4000\AA.  
We find: (i) our data allow for a robust merger fraction to be provided for a single redshift bin, which we center at $z=0.6$. 
(ii) Merger fractions at that $z$ have lower values than previous determinations: $f_{\rm m}^{\rm mph} = 0.045^{+0.014}_{-0.011}$ for $M_B \leq -20$ galaxies, and $f_{\rm m}^{\rm mph} = 0.031^{+0.013}_{-0.009}$ for $M_{K_{\rm s}} \leq -23.5$ galaxies. And,  
(iii) failure to address the effects of the large observational errors leads to overestimating $f_{\rm m}^{\rm mph}$ by factors of 10\%-60\%.  

Combining our results with those on other $B$-band selected samples, and parameterizing the merger fraction evolution as $f_{\rm m}^{\rm mph}(z) $ = $f_{\rm m}^{\rm mph}(0)$ $(1+z)^m$, we obtain that $m = 2.9 \pm 0.8$, and $f_{\rm m}^{\rm mph}(0) = 0.012 \pm 0.004$. For an assumed merger time-scale between 0.35-0.6 Gyr, these values imply that only 20\%-35\% of present day $M_B \leq -20$ galaxies have undergone a disk-disk major merger since $z \sim 1$

Assuming a $K_{\rm s}$-band mass-to-light ratio not varying with luminosity, we infer that the merger rate of galaxies with stellar mass $M_{\star} \gtrsim 3.5 \times 10^{10}\ M_{\odot}$ is $\Re_{\rm m} = 1.6^{+0.9}_{-0.6} \times 10^{-4}\ {\rm Mpc^{-3}}\ {\rm Gyr^{-1}}$ at $z=0.6$. When we compare with previous studies at similar redshifts, we find that the merger rate decreases when mass increases.
\end{abstract}


\setcounter{footnote}{0}

\section{INTRODUCTION}
Current $\Lambda$-cold dark matter simulations show that hierarchical halo mergers explain the build-up of dark matter structures in the Universe \citep{blumenthal84,springel05m}. While the merger history puts difficulties to the formation of disk galaxies \citep{vandenbosch01,abadi03,donghia04}, it is generally agreed that mergers are important in the formation of massive early-type galaxies \citep[e.g.,][]{delucia06}.  
Simulations suggest that gas-rich mergers can produce spheroidal systems \citep[e.g.,][]{naab03, bournaud05, hopkins08sim}, while dissipationless spheroidal mergers explain more massive spheroids \citep[e.g.,][]{gongarcia03, gongarcia05,naab06}. Mergers may also play a role in disk galaxy evolution: \citet{lotz08t} N-body simulations show that gas-rich major mergers can produce disk systems, while \citet{eliche06sim} show that minor mergers contribute to bulge growth in disk galaxies. Observationally, the size evolution of massive galaxies with redshift \citep{trujillo07,vandokkum08} and the luminosity density evolution of red sequence galaxies since $z \sim 1$ \citep{bell04,faber07} rule out the passive evolution hypothesis and suggest galaxy mergers as an important process in galaxy formation.

Despite their importance, the merger rate ($\Re_{\rm m}$, number of mergers per comoving volume and time), the merger fraction ($f_{\rm m}$, fraction of mergers in a given sample), and their evolution with $z$, are poorly constrained observationally. Merger fractions may be estimated from statistics of close pairs, or from statistics of geometrically-distorted galaxies.  
Working with close pairs \citep[e.g.,][]{patton00, patton02, lin04, lin08, depropris05, depropris07} gives useful information on the progenitors of the merger, such as their morphologies, their mass ratio, or their relative color. On the other hand, these studies need spectroscopic samples to avoid contamination by projection, which often leads to small sample sizes, although photometric redshift samples are starting to be used for this purpose \citep[e.g.,][]{bell06, kar07, ryan08, hsieh08}.

Methods based on morphological distortions, referred to in this paper as morphological merger fraction determinations, exploit the fact that, in the final stages of a disk-disk merger, the merger remnant is highly distorted, e.g., with high asymmetries, tidal tails, or double nuclei (\citealt{conselice03ff,conselice08,cassata05, kamp07,bridge07}; \citealt[][hereafter L08]{lotz08}). The identification of distorted sources may be done by eye \citep[e.g.,][]{kamp07}, or by automatic morphological indices, such as the asymmetry index A \citep{abraham96,conselice00}, and the $G$ and $M_{20}$ indices \citep{lotz04}. When using automatic methods, the uncertainties in the morphological indices and in the photometric redshifts, coupled to the fact that distorted galaxies represent a small fraction of the total, lead to objects statistically 'spilling over' from the most populated to the less populated bins, both in asymmetry and in photometric redshift.  These effects were studied by \citet[][hereafter LGB08]{clsj08}, who used maximum likelihood (ML) techniques to quantify the errors, and to provide unbiased determinations of the merger fractions. These authors conclude that the use of classical, straight histograms to compute merger fraction evolution can easily lead to overestimating merger fractions by $\sim$50\%. Such errors lead to overestimating the importance of mergers in galactic evolution models, e.g., to reproduce galaxy number counts, or when comparing observational results with $\Lambda$CDM predictions.

In this work we measure the morphological merger fraction at intermediate redshift ($z~\sim~0.6$) using asymmetry index diagnostics. Previous works find a merger fraction at that redshift of $\sim 0.07$ \citep{conselice03ff, lotz08}, although higher (0.09, \citealt{cassata05}; 0.16, \citealt{bridge07}), and lower (0.02, \citealt{kamp07}) values have been reported. We apply the ML method developed in LGB08 to determine to what degree previous merger fraction determinations are affected by redshift and asymmetry measurement errors.

The paper is structured as follows. In $\S$~\ref{data} we summarize the GOYA data used in this paper, while in $\S$~\ref{asy} we describe the asymmetry index calculations and its variation with redshift. In $\S$~\ref{metodo}, we review the ML method developed and tested in LBC08. The merger fraction values are presented in $\S$~\ref{results}, and we compare our results with those by other authors in $\S$~\ref{discusion}. Our conclusions are presented in $\S$~\ref{conclusion}. We use $H_0 = 70\ \rm{Km\ s^{-1}\ Mpc^{-1}}$, $\Omega_{M} = 0.3$, and $\Omega_{\Lambda} = 0.7$ throughout this paper. All magnitudes are in Vega system, unless noted otherwise. 

\setcounter{footnote}{0}

\section{DATA}\label{data}
We work with images and catalogs from the GOYA photometric survey, an imaging survey in preparation for the GOYA\footnote{http://www.astro.ufl.edu/GOYA/home.html} NIR spectroscopic survey with GTC/EMIR \citep{guzman03}.  
For this paper, we focus on the Groth strip (GS) field, which is covered in six broadband filters ($U, B, V, I, J, K_{\rm s}$) over a common area of 155 square arcmin.  
The area covered is that of the original Groth strip survey with HST/WFPC2 \citep{groth94}, centered on $\alpha = 14^{\rm h}16^{\rm m}38^{\rm s}_{\cdot}8$ and $\delta = 52^{\circ}16^{\prime}52^{\prime\prime}$ (J2000.0). HST/WFPC2 imaging, which will provide the data for our asymmetry measurements, have been extensively described elsewhere \citep[e.g.][]{ratnatunga95, ratnatunga99, simard02}. Exposure times for the images used here were 2800 s in $F606W$ ($V_{606}$) and 4400 s in $F814W$ ($i_{814}$), and a typical depth of $i_{814} = 25.0$ is reached.

$K_{\rm s}$-band imaging is described in \citet{cristobal03}.  The Groth strip was covered with 11 pointings of the WHT/INGRID camera, with a pixel scale of 0.24 arcsec and seeing ranging from 0.6\arcsec\ to 1.1\arcsec\ FWHM.  With typical exposure times of 5700 s, the median 3-$\sigma$ depth is 20.5 mag.  Imaging in the $J$ band was similarly carried out with the WHT/INGRID camera.  Exposure times were 1800s, leading to depths of 21.8 mag.  

$U$ and $B$-band imaging are described in \citet{eliche06}. The entire Groth strip was covered with a single pointing of the INT/WFC.  Integration times were 14,400 s in $U$ and 10,300 s in $B$, which led to $\sim$3-$\sigma$ depths of 24.8 mag in $U$ and 25.5 mag in $B$. The FWHM of the images was 1.3\arcsec.

\subsection{The GOYA-GS Catalog}
\begin{figure}[t]
\plotone{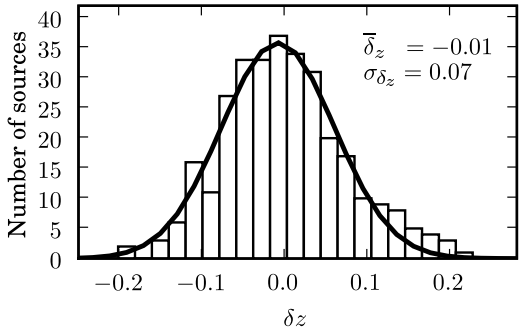}
\caption{Histogram of the variable $\delta_z$ (see text for definition). The black solid line is the best fit of the histogram to a Gaussian with $\overline{\delta}_z = -0.01$ and $\sigma_{\delta_z} = 0.07$.}
\label{dz}
\end{figure}

The parent catalog for the present study is a $K_{\rm s}$-selected catalog comprising 2450 sources, for which photometry is provided in six bands, $U$ through $K_{\rm s}$. The limiting magnitude of the catalog, as derived from simulations with synthetic sources, is $K_{\rm s} \sim 20.51$ (50\% detection efficiency). Multi-band photometry comes from images with matched PSF to 1.3\arcsec\ of FWHM.  
The catalog contains DEEP3\footnote{http://deep.berkeley.edu/DR3/} spectroscopic redshifts ($z_{\rm spec}$) for $\sim 600$ sources and photometric redshifts ($z_{\rm phot}$) for all.  
The latter were obtained with \textsc{HyperZ} \citep{bolzonella00}, that also provided the 68\% confidence interval for each $z_{\rm phot}$. However, the probability distributions that describe the $z_{\rm phot}$'s are not symmetric, present non-analytic shapes and can have double peaks, while our methodology needs that errors in $z_{\rm phot}$ be Gaussian (\S~\ref{metodo}, LGB08). In addition, we find that the confidence intervals given by \textsc{HyperZ} do not correlate with the differences between $z_{\rm spec}$'s and $z_{\rm phot}$'s, a result also noted by \citet{oyaizu08}. Because of this, we used $\sigma_{z_{\rm phot}} = \sigma_{\delta_z} (1+z_{\rm phot})$ as $z_{\rm phot}$ error, where $\sigma_{\delta_z}$ is the standard deviation in the distribution of the variable $\delta_z \equiv (z_{\rm phot} - z_{\rm spec}) / ({1 + z_{\rm phot}})$, that is well described by a Gaussian with $\overline{\delta}_z = -0.01$ and $\sigma_{\delta_z} = 0.07$ (Fig.~\ref{dz}). This procedure assigned the same error to sources with equal $z_{\rm phot}$, but it is statistically representative of our sample and ensures the Gaussianity of $z_{\rm phot}$ errors in the merger fraction determination (\S~\ref{metodo}), while 68\% confidence intervals from \textsc{HyperZ} do not.

\textsc{HyperZ} also yields the most probable SED of the source, that is used for degradation of the sources in the asymmetry calculation process (see \S\S~\ref{homoa} and \ref{avsz}) and for computing absolute magnitudes, necessary for doing the final selection of our sample (see the next section).

\subsection{Galaxy Samples}\label{bkssample}
We define two samples for our morphological analysis, selected in $B$- and $K_{\rm s}$-band absolute magnitude, respectively.  The $B$-selected sample allows for comparisons with other studies in the literature, which often select their samples in that visual band. Comparisons are carried out in \S~\ref{mphdisc}. On the other hand, absolute $K_{\rm s}$ magnitude is a good tracer of the stellar mass of the galaxy \citep{bell01,drory04}. This makes the galaxy selection less dependent on the instantaneous star formation, giving us a more nearly mass-selected determination of the merger fraction. 

The redshift range for our samples is determined as follows. 
The highest redshift at which we can use asymmetry as a reliable morphological indicator without perform a morphological K-correction is $z_{\rm up} = 0.85$ (see $\S$~\ref{zrange}). In addition, because the ML method used in the merger fraction determination (see $\S$~\ref{metodo}) takes into account the experimental errors, we must include in the samples not only the sources with $z_i < z_{\rm up}$, but also sources with $z_i-2\sigma_i < z_{\rm up}$ in order to ensure completeness. Because of this, the maximum redshift in our samples, $z_{\rm max}$, must fulfill the condition $z_{\rm max} - 2\sigma_{\delta_z}(1 + z_{\rm max}) = 0.85$, which yields $z_{\rm max} \sim 1.15$. We take as minimum redshift in our study $z_{\rm min} = 0.2$ because of the lack of sources at lower redshifts. This yields $z_{\rm down} = z_{\rm min}+2\sigma_{\delta_z}(1 + z_{\rm min}) \sim 0.35$ to ensure completeness and good statistics.

\begin{figure}[t]
\plotone{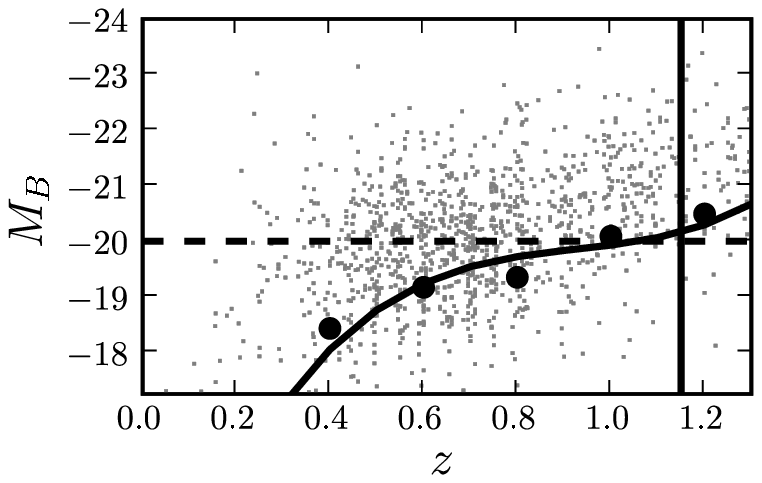}
\plotone{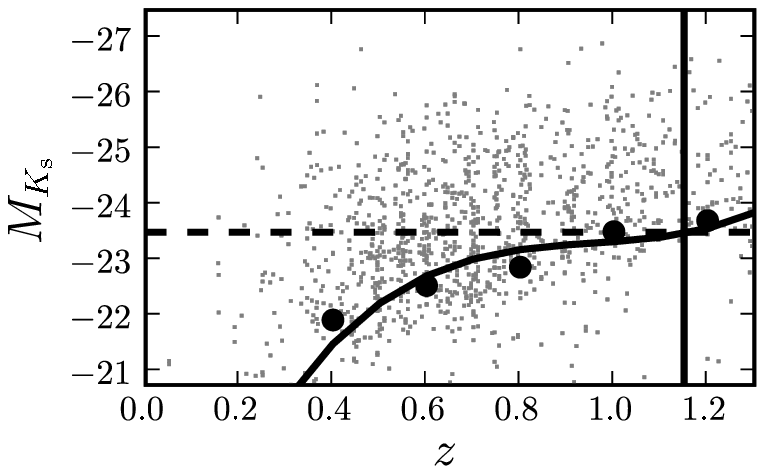}
\caption{$M_B$ ({\it upper panel}) and $M_{K_{\rm s}}$ ({\it lower panel}) as function of redshift ({\it gray dots}). In both panels black dots mark the limiting magnitude at different redshifts. The black solid curves are the least-squares fits of the completeness points by a third degree polynomial. The horizontal black dashed lines are the magnitude cuts that we used to select our samples: $M_B \leq -20$ ({\it upper}) and $M_{K_{\rm s}} \leq -23.5$ ({\it lower}).}\label{mbkvsz}
\end{figure}

The limiting absolute magnitude of the samples is obtained by calculating the third quartile of the $M_B$ and $M_{K_{\rm s}}$ redshift distributions in various redshift bins.  We show those as filled circles in Figure~\ref{mbkvsz}, for $M_B$ ({\it upper panel}) and $M_{K_{\rm s}}$ ({\it lower panel}). The black solid curves are the least-squares fits to these points by a third degree polynomial. To be complete up to $z_{\rm max} = 1.15$ (indicated with {\it vertical solid lines}), the samples need to be restricted to $M_B \leq -20$ and $M_{K_{\rm s}} \leq -23.5$. These limits are indicated as horizontal dashed lines in both panels.  The $B-K_{\rm s}$ color distribution of the two samples peaks at $B-K_{\rm s} = 3.6$, consistent with the difference between both selection cuts, which is 3.5 magnitudes. This selection yields 567 sources with $M_B \leq -20$ and $0 < z < 1.15$, and 505 with $M_{K_{\rm s}} \leq -23.5$ and $0 < z < 1.15$. We study the difference between both catalogs in $\S$~\ref{bvsk}.

\section{ASYMMETRY INDEX}\label{asy}
The automatic asymmetry index $A$ is one of the CAS morphological indices \citep[][hereafter C03]{conselice03}. It is defined as:
\begin{equation}
A = \frac{\sum |I_0 - I_{180}|}{\sum |I_0|} - \frac{\sum |B_0 - B_{180}|}{\sum |I_0|},\label{A}
\end{equation}
where $I_0$ and $B_0$ are the original galaxy and background images, $I_{180}$ and $B_{180}$ are the original galaxy and background images rotated by 180 degrees, and the sum spans all the pixels of the galaxy and background images. The index increases with the deviation of the galaxy image from point-symmetry.  The background terms in equation~(\ref{A}) are measured on a region of the frame free from known sources; it takes into account the level of asymmetry expected from the noise distribution in the sky pixels. Further details on the asymmetry calculation are given in \citet[][]{conselice00}. We use the $A$ index to identify recent merger systems which are very distorted. On the basis of asymmetry measurements on images of nearby merger remnants, previous merger fraction determinations have taken a system to be a major merger remnant if its asymmetry index is $A > A_{\rm m}$, with $A_{\rm m} = 0.35$ (C03). Note that this criterion applies to disk-disk mergers only; spheroid-dominated mergers suffer much weaker morphological distortions, hence they are missed by the asymmetry criterion just described.  For high-redshift samples, the determination of $A$ needs to be done on HST images to ensure high spatial resolution.  In our case, we work with $V_{606}$ and $i_{814}$ bands. To increase the signal-to-noise we determined the asymmetry index $A_0$ of each source in the image $V_{606} + i_{814}$.

\subsection{Pass-band Restrictions to the Redshift Range}\label{zrange}
\begin{figure}[t]
\plotone{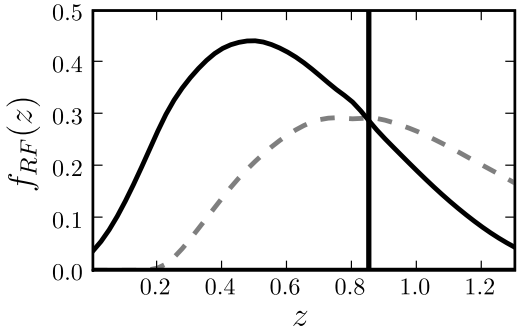}
\caption{Variation of $f_{RF}(z)$ with redshift. Black solid line is for $B$ filter as reference, while gray dashed line is for $U$ filter. The vertical black solid line indicates $z_{\rm up}  = 0.85$.}
\label{fz}
\end{figure}

Galaxy morphology depends on the band of observation \citep[e.g.][]{kuchinski00, taylor07}.  In particular, when galaxies contain both old and young populations, morphologies may change very significantly at both sides of the Balmer/4000\AA\ break. The asymmetry index limit $A_{\rm m} = 0.35$ was established in the rest-frame $B$-band (see C03).  When dealing with galaxies over a range of redshifts, in order to avoid systematic passband biases with redshift, one needs to apply a so-called morphological K-correction by performing the asymmetry measurements in a band as close as possible to rest-frame $B$ \citep[e.g.,][]{cassata05}.  While corrections have been attempted for obtaining asymmetries in rest-frame $B$ from asymmetry measurements in rest-frame $U$ \citep{conselice08}, in the present study we stay within the redshift range where our images sample rest-frame $B$.  
To determine the redshift ranges over which rest-frame $B$-band or $U$-band dominates the flux in the observational $V_{606}+i_{814}$ filter, we define the function
\begin{equation}
f_{\rm RF}(z) = \frac{\int_0^{\infty}P_{Vi}(\lambda/(1+z)) P_{\rm RF}(\lambda){\rm d}\lambda}{\int_0^{\infty}P_{\rm RF}(\lambda){\rm d}\lambda}, 
\end{equation}
where $P_{\rm RF}$ and $P_{Vi}$ are the transmission curves of the rest-frame reference filter and the $V_{606}+i_{814}$ filter, respectively. In Figure~\ref{fz} we show the function $f_{B}(z)$ ({\it black curve}), and $f_{U}(z)$ ({\it gray dashed curve}). The redshift in which the $U$-band starts to dominate the flux in the observed $V_{606}+i_{814}$ filter is $z_{\rm up}=0.85$ ({\it vertical black solid line}). We take this redshift as the upper limit for our study.

\subsection{Asymmetries at a Reference Redshift}\label{homoa}
The asymmetry index measured on survey images varies systematically with the source redshift, due to the loss of spatial resolution and of source flux with $z$. The net result of such loss of information is a systematic decrease with $z$ of the measured asymmetry.  Several papers have attempted to quantify these effects by degrading the image spatial resolution and flux to simulate the appearance a given galaxy would have at different redshifts in a given survey. 
\citet{conselice03ff, conselice08}; and \citet{cassata05} degraded a few local galaxies to higher redshifts, and found that indeed asymmetries decrease with $z$. \citet{conselice03ff} also noted that this descent depends on image depth, and that luminous galaxies are less affected. In addition, \citet{conselice05} show that irregular galaxies (high asymmetry) are more affected than ellipticals (low asymmetry). 
A zeroth-order correction for such biases was implemented by \citet{conselice03ff,conselice08} who applied a correction term $\Delta A_z$ defined as the difference between the asymmetry of local galaxies measured in the original images and the asymmetry of the same galaxies in the images degraded to redshift $z$. Their final, corrected asymmetries are $A_{\rm f} = A_0 + \Delta A_z$, where $A_0$ is the asymmetry measured in the original images. With these corrections, all the galaxies have their asymmetry referred to $z = 0$, and the local merger criterion $A > A_{\rm m} = 0.35$ is used to flag merger remnants.

We improve on the above procedure by computing a correction term individually for each source in the catalog.  Also, rather than attempting to recover $z=0$ values for $A$, we degrade each of the galaxy images to redshift $z_{\rm d} = 0.75$; we then obtain our final asymmetry values $A_{\rm f}$ directly from the degraded images.  With this procedure, we take into account that each galaxy is affected differently by the degradation, e.g., the asymmetry of a low luminosity irregular galaxy dramatically decreases with redshift, while a luminous elliptical is slightly affected.  We choose $z_{\rm d} = 0.75$ as our reference redshift because a source at this (photometric) redshift has $z_{\rm d} + \sigma_{z_{\rm d}} \sim z_{\rm up} = 0.85$, that is, the probability that our galaxy belongs to the range of interest is $\sim 85$\%. 
Because we will work with asymmetries reduced to $z_{\rm d} = 0.75$, we cannot apply the local merger criterion $A > A_{\rm m} = 0.35$.  We redefine this criterion in $\S$~\ref{newAm}.

Only $\sim$26\% of the sources in the catalog have spectroscopic redshifts, hence redshift information, coming primarily from photometric redshifts (\S~\ref{data}), has large uncertainties.    
To account for the redshift uncertainty when deriving the asymmetries at $z_{\rm d}=0.75$, we start from three different initial redshifts for each photometric source, $z_{\rm phot}^{-} = z_{\rm phot} - \sigma_{z_{\rm phot}}$, $z_{\rm phot}$, and $z_{\rm phot}^{+} = z_{\rm phot} + \sigma_{z_{\rm phot}}$, and degrade the image from these three redshifts to $z_{\rm d} = 0.75$. Then, we perform a weighted average of the three asymmetry values, such that 
\begin{eqnarray}
A_{\rm f}&=&0.16A_{0.75}(z_{\rm phot}^{-}) + 0.16A_{0.75}(z_{\rm phot}^{+})\nonumber\\ 
&+& 0.68A_{0.75}(z_{\rm phot}),\label{prom}
\end{eqnarray}
where $A_{0.75}(z)$ denotes the asymmetry measured in the image degraded from $z$ to $z_{\rm d} = 0.75$. When a spectroscopic redshift is available, the final asymmetry is simply $A_{\rm f} = A_{0.75}(z_{\rm spec})$. We do not apply any degradation to sources with $z > 0.75$, that is, we assume that $A_{0.75}(z > 0.75) = A_0$. Whenever a source is not detected  after degradation, we remove it from the sample in spectroscopic redshift cases, and we do not use it in equation~(\ref{prom}) in photometric redshift cases.

To obtain the error of the asymmetry, denoted $\sigma_{A_{\rm f}}$, we average, in photometric redshift cases, the uncertainties of the three asymmetries following equation~(\ref{prom}), and add the result in quadrature to the $rms$ of the three asymmetry values. The first term accounts for the signal-to-noise error in the asymmetry value, while the second term is only important when differences between the three asymmetry values can not be explained by the signal-to-noise first term.
In the spectroscopic case we take as $\sigma_{A_{\rm f}}$ the uncertainty of the asymmetry $A_{0.75}(z_{\rm spec})$.

The degradation of the images was done with \textsc{cosmoshift} \citep{balcells03}, which performs repixelation, PSF change,  flux decrease, and K-correction over the sky-subtracted source image. The K-correction for each source is computed through integrals of the best-fit SED obtained as output of \textsc{HyperZ} in the $z_{\rm phot}$ determination. The last \textsc{cosmoshift} step is the addition of a random Poisson sky noise to the degraded source model. As a result of this last step, two \textsc{cosmoshift} degradations of the same source will yield different asymmetry determinations. We take the asymmetry of each degraded source, $A_{0.75}(z)$, to be the median of asymmetry measurements on 10 independent degradations of the original source image from $z$ to $z_{\rm d} = 0.75$. Note that, in photometric redshift cases, each final asymmetry value comes from three previous asymmetries (eq.~[\ref{prom}]), so each $A_{\rm f}$ determination involves 30 asymmetry calculations. In all the cases the uncertainty in $A_{0.75}(z)$ is the median of the 10 individual asymmetry errors. 

To check that the different final asymmetry determinations for sources with photometric and spectroscopic redshifts do not bias the asymmetry values, we compare the $A_{\rm f}$ of the 56 sources with $z_{\rm spec}$, $M_B \leq -20$, and $0.35 \leq z_{\rm spec} < 0.75$, denoted $A_{\rm f}({\rm spec})$, with the final asymmetries obtained from their corresponding $z_{\rm phot}$ and Equation~\ref{prom}, denoted $A_{\rm f}({\rm phot})$. The difference $A_{\rm f}({\rm spec}) - A_{\rm f}({\rm phot})$ has an $rms = 0.025$, lower than the typical error in $A_f$ ($\overline{\sigma_{A_f}} \sim 0.04$). In fact, 90\% of the sources have their $A_{\rm f}({\rm phot})$ inside $\pm \sigma_{A_f}({\rm spec})$ and all sources inside $\pm 2\sigma_{A_f}({\rm spec})$. Therefore, we conclude that the $A_{\rm f}$ measured in sources with $z_{\rm phot}$ are equivalent to the $A_{\rm f}$ measured in sources with $z_{\rm spec}$.

The asymmetries $A_{\rm f}$ referred to $z_{\rm d}=0.75$ provide a homogeneous asymmetry set that permits consistent morphological studies in the GS field. In $\S$~\ref{mlimpor} we discuss the effects that the usage of degraded or non-degraded asymmetries has on the merger fraction determination.

\subsection{Asymmetry Trends with Redshift}\label{avsz}
For a sample of galaxies over a range of redshifts, the statistical change of the measured asymmetries with $z$ is the combined effect of loss of information (as shown in the previous section) and changes of the galaxy populations.  In contrast, the $z$ evolution of $A_{\rm f}$ reflects changes in the galaxy population alone, given that morphological information in the images used to determine  $A_{\rm f}$ is homogeneous for the sample.  We show here that the $z$ trends of $A_0$ and  $A_{\rm f}$ are quite different.  

\begin{figure}[t]
\plotone{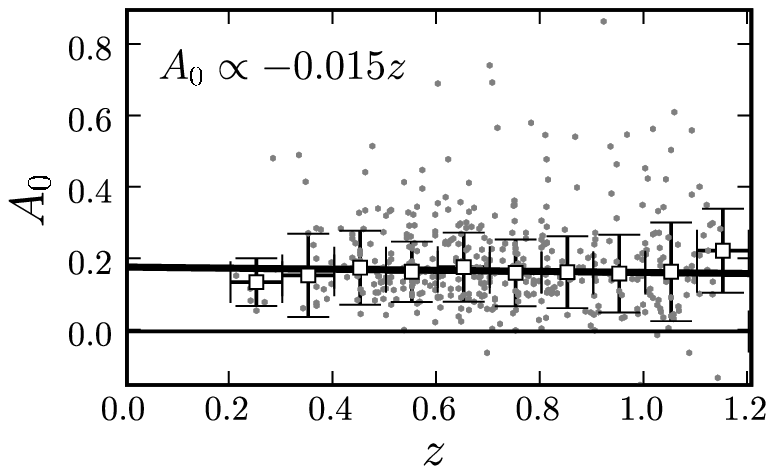}
\plotone{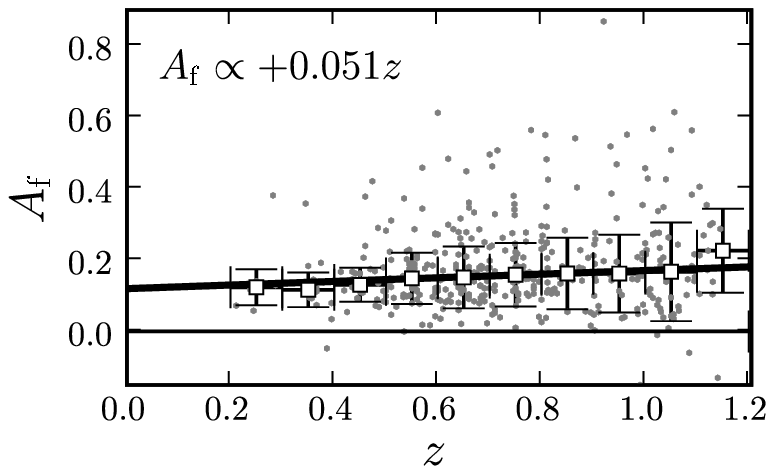}
\caption{Asymmetry vs. redshift in the $M_B \leq -20$ sample ({\it gray dots} in both panels). {\it Upper}: asymmetries of the sources measured on the original images. {\it Lower}: asymmetries of the sources measured on images degraded to $z_{\rm d} = 0.75$. White squares in both panels are the mean asymmetries in 0.1 redshift bins. The black solid line is the linear fit to the mean asymmetries in the [0.5,0.8) redshift interval.}\label{asyevol}
\end{figure}

In the upper panel of Figure~\ref{asyevol} we show the variation of $A_0$ with redshift in the $M_B \leq -20$ sample, while in lower panel we see the variation of $A_{\rm f}$ for the same sample. In both panels, white squares are the median asymmetries in $\Delta(z) = 0.1$ redshift bins, and the black solid line is the linear least-squares fit to the $0.5 \leq z < 0.8$ points. In the $A_0$ case the slope of the fit is negative, $A_0 \propto -0.015 z$, while in the $A_{\rm f}$ case the slope is positive, $A_{\rm f} \propto 0.051 z$. In the first case, the negative slope reflects that the loss of information with redshift (negative effect on $A$) dominates over genuine population variations (positive effect, because at higher redshift galaxies are more asymmetric, e.g., \citealt{cassata05,conselice05}). In the second case, we have removed the loss of information term, so we only see population effects. We take as degradation rate, denoted $\delta_A$, the difference between both slopes, that yields $\Delta A_B = \delta_A \Delta z = -0.066 \Delta z$. In the $K_{\rm s}$-limited sample we follow the same procedure and obtain $\Delta A_{K_{\rm s}} = -0.060 \Delta z$.

\subsection{Adjusting the Asymmetry-based Merger Criterion}\label{newAm}
How do the trends shown in $\S$~\ref{avsz} affect the merger criterion $A_{\rm m} = 0.35$? This value was chosen in C03 to select the high asymmetry tail of the local asymmetry distribution: most of the local galaxies with $A > 0.35$ are merger systems.  Note however that only $\sim 50$\% of all local merger systems have $A > 0.35$ (C03, Fig. 9). If asymmetry systematically decreases with redshift, the local merger criterion needs to descrease as well to pick up the same distorted sources of the sample. This idea is supported by \citet{kamp07}, who visually compare distorted galaxies at $z \sim 0.7$ with their asymmetry values. They find that a third to a half of the galaxies with $A > 0.20$ are mergers, while 60\% of merger systems have $A > 0.20$ \citep[][their Fig. 5]{kamp07}.  Hence, although not all the sources with $A > 0.20$ are merger systems, the number of galaxies with $A > 0.20$ are statically representative of the total number of mergers. We assume that the local merger criterion evolves with redshift as $A_{\rm m}(z) = A_{\rm m}(0) + \delta_A z = 0.35 + \delta_A z$. With the two different values of $\delta_A$ obtained previously, we infer that $A_{\rm m}(0.75) \sim  0.30$, and we use this merger criterion in the following. This variation is particular of our samples and HST/WFPC2 images: deeper images, and images with different angular resolution may show different trends.

\section{MAXIMUM-LIKELIHOOD MERGER FRACTION DETERMINATION}\label{metodo}
Following \citet[][]{conselice06}, the merger fraction by morphological criteria is
\begin{equation}\label{fmg}
f^{\rm mph} = \frac{\kappa N_{\rm m}}{N_{\rm tot} + (\kappa - 1) N_{\rm m}},
\end{equation}
where $N_{\rm m}$ is the number of the distorted sources in the sample with $A > A_{\rm m}$, and $N_{\rm tot}$ is the total number of sources in the sample. If $\kappa \geq 2$ we obtain the galaxy merger fraction, $f_{\rm gm}^{\rm mph}$, and $\kappa$ represents the average number of galaxies that merged to produce one distorted remnant. If $\kappa = 1$ we obtain the merger fraction, $f_{\rm m}^{\rm mph}$: the number of merger events in the sample. We will use $\kappa = 1$ throughout this paper.

The steps that we follow to obtain the merger fraction are described in detail in LGB08. In this section we review the main steps. If we define a bidimensional histogram in the redshift--asymmetry space and normalize this histogram to unity, we obtain a bidimensional probability distribution defined by the probability of having one source in bin $[z_k, z_{k+1}) \cap [A_l, A_{l+1})$, defined as $p_{kl}$, where index $k$ spans the redshift bins of size $\Delta z$, and the index $l$ spans the asymmetry bins of size $\Delta A$. We consider only two asymmetry bins split at $A_{\rm m}$, such that  the probabilities $p_{k1}$ describe highly distorted galaxies (i.e., merger systems), while the probabilities $p_{k0}$ describe normal galaxies. With those definitions, the merger fraction in the redshift interval $[z_k, z_{k+1})$ becomes
\begin{equation}\label{ffpkl}
f_{{\rm m},k}^{\rm mph} = \frac{p_{k1}}{p_{k0}+p_{k1}}.
\end{equation}
In LGB08 we developed a ML method that yields the most probable values of $p_{kl}$ taking into account not only the $z$ and $A$ values, but also their experimental errors. The method is based on the minimization of the joint likelihood function, that in our case is
\begin{displaymath}
L(z_{i},A_{i}|p^{\prime}_{kl},\sigma_{z_{i}},\sigma_{A_{i}})\ \ \ \ \ \ \ \ \ \ \ \ \ \ \ \ \ \ \ \ \ 
\end{displaymath}
\begin{equation}
= \sum_i \biggr[ \ln \bigg\{ \sum_k\sum_l \frac{{\rm e}^{p'_{kl}}}{4}{\rm ERF}(z,i,k){\rm ERF}(A,i,l) \bigg\}\biggr]\label{MLfunc},
\end{equation}
where
\begin{equation}
{\rm ERF}(\eta,i,k) \equiv {\rm erf}\bigg(\frac{\eta_{i} - \eta_{k+1}}{\sqrt{2} \sigma_{\eta_{i}}}\bigg) - {\rm erf}\bigg(\frac{\eta_{i} - \eta_{k}}{\sqrt{2} \sigma_{\eta_{i}}}\bigg),\label{ERF}
\end{equation}
${\rm erf}(x)$ is the error function, $z_i$ and $A_i$ are the redshift and asymmetry values of source $i$, respectively, $\sigma_{z_i}$ and $\sigma_{A_i}$ are the observational errors in redshift and asymmetry of source $i$, respectively, and the new variables $p'_{kl} \equiv \ln (p_{kl})$ are chosen to avoid negative probabilities. Equation~(\ref{MLfunc}) was obtained by assuming that the real distribution of galaxies in the redshift--asymmetry space is described by a bidimensional distribution $p_{kl} = \exp(p'_{kl})$ and that the experimental errors are Gaussian (see LGB08 for details). Note that changing variables to $p'_{kl} = \ln(p_{kl})$, equation~(\ref{ffpkl}) becomes
\begin{equation}\label{ffpklp}
f_{{\rm m},k}^{\rm mph} = \frac{{\rm e}^{p'_{k1}}}{{\rm e}^{p'_{k0}}+{\rm e}^{p'_{k1}}}.
\end{equation}
LGB08 show, using synthetic catalogs, that the experimental errors tend to smooth an initial bidimensional distribution described by $p_{kl}$, due to spill-over of sources to neighboring bins.  This leads to a $\sim10-30$\% overestimate of the galaxy merger fraction in typical observational cases. L08 find similar trends in their study of morphological merger fraction based on the $M_{20}$ and $G$ indices.
LGB08 additionally show that, thanks to the use of the ML method, we accurately recover the initial bidimensional distribution: the input and ML method merger fraction difference is $\sim 1$\% even when experimental errors are similar to the bin size. That is, the ML results are not biased by the spill-over of sources to neighboring bins.

We obtain the morphological merger fraction by applying equation~(\ref{ffpklp}) using the probabilities $p'_{kl}$ recovered by the ML method.
In addition, the ML method provides an estimate of the 68\% confidence intervals of the probabilities $p'_{kl}$, that we use to obtain the $f_{{\rm m},k}^{\rm mph}$ 68\% confidence interval, denoted $[\sigma^{-}_{f_{{\rm m},k}^{\rm mph}}, \sigma^{+}_{f_{{\rm m},k}^{\rm mph}}]$. This interval is asymmetric because $f_{{\rm m},k}^{\rm mph}$ is described by a log-normal distribution due to the calculation process (see LGB08 for details). Note that, in LGB08, $\kappa = 2$ is used in equation~(\ref{fmg}), but the method is valid for any $\kappa$ value.

\subsection{Simulations with synthetic catalogs}\label{simu}
LGB08 show that the reliability of the ML method depends on factors such as the number of sources in the catalog, the mean experimental errors relative to the bin sizes, and the values of $p_{kl}$ (bins with lower probabilities are more difficult to recover). LGB08 also show that the probability distributions of $p'_{kl}$ must be Gaussian to ensure the reliability of the method. We study the shape of probability distributions of $p'_{kl}$ by performing simulations with synthetic catalogs.  

We characterize the experimental catalogs with several parameters, which we use as input for the synthetic catalogs. These parameters are: the number of sources ($n$), the fraction of spectroscopic redshift sources in each redshift bin ($f_{k, \rm{spec}}$), and the mean and the dispersion of $A_{\rm f}$ errors in each redshift bin ($\overline{\sigma_A}_{,k}$, and $\sigma_{\sigma_{A, k}}$). Additionally, each experimental catalog have associated the $p'_{kl}$ probabilities that we obtain previously applying the ML method. To obtain these $p'_{kl}$ we fix the asymmetry bin size $\Delta A = 0.5$, and only vary the redshift bin size $\Delta z$, that is, $\Delta z$ is the only free parameter in this study.

With the previous input parameters we create a synthetic catalog as follows: first we take $n$ random sources distributed in redshift and asymmetry space following a bidimensional distribution defined by the probabilities $p_{kl} = \exp(p'_{kl})$. This process yields the input values of $z$ and $A$, $z_{{\rm in},i}$ and $A_{{\rm in},i}$, of the $n$ sources of our synthetic catalog, and the number of sources in each redshift bin, $n_{k}$. Next we apply the experimental redshift errors. For $n_{k} f_{k, \rm{spec}}$ sources in each redshift bin we assume that the simulated redshift value is equal to the input value, $z_{{\rm sim},i} = z_{{\rm in},i}$, and assign it a constant standard deviation $\sigma_{z_{{\rm sim},i}} = 0.001$. For the remaining $n_{k}(1 - f_{k, \rm{spec}})$ sources in each redshift bin, the process is more complicated: we obtain the $z_{{\rm sim},i}$ value as drawn for a Gaussian distribution with mean $z_{{\rm in},i}$ and standard deviation $\sigma_{z_{{\rm sim},i}} = 0.07(1+z_{{\rm in},i})$. The process to obtain the simulated asymmetry values $A_{{\rm sim},i}$ is similar: these as drawn for a Gaussian distribution with mean $A_{{\rm in},i}$ and standard deviation $\sigma_{A_{{\rm sim},i}}$. In this case, the value of $\sigma_{A_{{\rm sim},i}}$ is a positive value also drawn for a Gaussian distribution with mean $\overline{\sigma_A}_{,k}$ and standard deviation $\sigma_{\sigma_{A,k}}$, so it depends on the redshift $z_{{\rm in},i}$ of the source.

In order to characterize the probability distributions that describe the $p'_{kl}$, we generate a set of $N = 1000$ independent synthetic catalogs and apply the ML method to each catalog. We find that, due to the low number of highly asymmetric sources, the Gaussianity of the $p'_{kl}$ can only be ensured if we consider one redshift bin at $z_{\rm down} = 0.35 \leq z < z_{\rm up} = 0.85$ range (see $\S$~\ref{bkssample} for details about these limits), that is, $\Delta z = 0.5$. If we consider two redshift bins with $\Delta z = 0.25$, the probability distributions of the $p'_{k1}$ are non-Gaussian and we cannot ensure the reliability of the results.\\

\begin{deluxetable*}{lcccc} 
\tabletypesize{}
\tablecolumns{5} 
\tablewidth{0pc} 
\tablecaption{Catalog Sources with $z \in [0.35, 0.85)$\label{ffncat}}
\tablehead{ 
Sample selection & $n_{\rm class}$ & $n_{\rm class}(A_{\rm f} > 0.30)\tablenotemark{a}$ & $n_{\rm ML}\tablenotemark{b}$ & $n_{\rm ML}(A_{\rm f} > 0.30)\tablenotemark{c}$}
\startdata
$M_{B} \leq -20$    & 352 & 25 & 383.4 & 17.4 \\
$M_{K_{\rm s}} \leq -23.5$ & 313 & 14 & 348.6 & 10.8 \\
\enddata 
\tablenotetext{a}{Number of merger systems in the bin.}
\tablenotetext{b}{Number of sources in the bin given by ML method. See $\S$~\ref{results} for details.}
\tablenotetext{c}{Number of merger systems in the bin given by ML method. See $\S$~\ref{results} for details.}
\end{deluxetable*}

\section{RESULTS}\label{results}
On the basis of the arguments in $\S\S$~\ref{asy} and \ref{metodo}, we provide merger fractions for the redshift interval $z \in [0.35, 0.85)$. 
In Table~\ref{ffncat} we summarize the total number of sources, and the number of distorted sources ($A_{\rm f} > 0.30$), with $z \in [0.35, 0.85)$, for both classical counting and ML method. For the ML method, the number of sources is not an integer. Indeed, the ML method gives a statistical estimation of the probability $p_{kl} = \exp(p'_{kl})$ of finding one source in the redshift bin $k$ and in the asymmetry bin $l$, so the estimated number of galaxies in that bin, $n_{kl,{\rm ML}} = n_{\rm tot}p_{kl}\Delta z \Delta A$ (where $n_{\rm tot}$ is the total number of galaxies in the sample), is not necessarily an integer. Table~\ref{ffncat} shows that the number of distorted sources in the $K_{\rm s}$-band sample is lower that in $B$-band sample.  This result, which occurs for both classical and ML method determinations, is analyzed in more detail in $\S$~\ref{bvsk}. In addition, the number of distorted sources given by the ML method is lower than that coming from the classical determination, but the total number of sources in the bin is higher. This is due to the fact that most of the sources in the samples ($\sim70$\%) are in the range $z \in [0.35, 0.85)$, therefore more sources have spilled out of this bin due to redshift errors than viceversa.  

With the probabilities $p'_{kl}$ and their confidence intervals given by the ML method, we obtain the following merger fractions for our $B$ and $K_{\rm s}$ samples:
\begin{eqnarray}
f_{\rm m}^{\rm mph}(z = 0.6, M_B \leq -20) = 0.045^{+0.014}_{-0.011},\\
f_{\rm m}^{\rm mph}(z = 0.6, M_{K_{\rm s}} \leq -23.5) = 0.031^{+0.013}_{-0.009}.
\end{eqnarray}

In the next section we study the origin of the difference between both values, and we compare our results to other authors in $\S$~\ref{discusion}.

\subsection{Visible vs Near Infrared Merger Fractions}\label{bvsk}
\begin{figure}[t]
\plotone{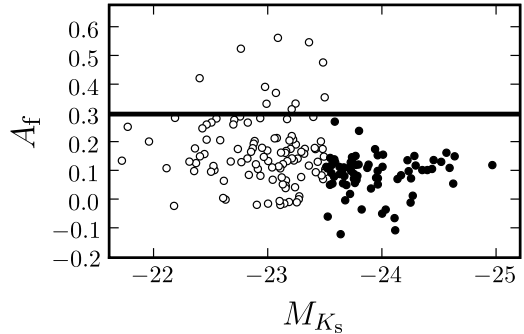}
\caption{Asymmetry in function of $M_{K_{\rm s}}$ for blue ({\it white dots}) and red ({\it black dots}) samples (see the text for details). The black solid line shows the merger criterion $A_{\rm m} = 0.30$.}\label{zabks}
\label{AvsZ}
\end{figure}

We found that the merger fraction at $z = 0.6$ in the $M_{K_{\rm s}}$ selected sample is a $\sim30$\% lower than in the $M_B$ selected sample. Such trend had previously been noted in pair studies \citep{bundy04,rawat08}. To understand the origin of this difference, we study the nature of the galaxies in the range $0.35 \leq z < 0.85$ that are \textit{not common} to the two samples. For this discussion, we shall refer to galaxies only selected in $B$/$K_{\rm s}$ samples as the blue/red samples. The blue/red samples comprise 109/72 sources. As expected, the blue sample comprises lower mass galaxies ($\langle M_{K_{\rm s}} \rangle_{\rm blue} = -23.1$ vs $\langle M_{K_{\rm s}} \rangle_{\rm red} = -23.8$), which are bluer than those from the red sample ($\langle [B-K_{\rm s}] \rangle_{\rm blue} = 2.6$ vs $\langle [B-K_{\rm s}] \rangle_{\rm red} = 4.4$).  We now show that the two samples have very distinct asymmetry distributions.  
In Figure~\ref{zabks} we plot the blue ({\it open symbols}) and red ({\it filled symbols}) samples in the $A_{\rm f}$--$M_{K_{\rm s}}$ plane. It is clear that asymmetries in the red sample are low, $\langle A_{\rm m} \rangle_{\rm red} = 0.10$, with only 2 sources with $A_{\rm f} > 0.2$ and none with $A_{\rm f} > 0.3$ ({\it black solid line}). In contrast, the mean asymmetry of the blue sample is $\langle A_{\rm m} \rangle_{\rm blue} = 0.15$, with 11 sources with $A_{\rm f} > 0.3$ ($\sim10$\% of the blue sample). 
This result suggests that: (i) an important fraction of the $B$-band high asymmetry sources are low-mass disk-disk major merger systems that, due to merger-triggered star formation, have their $B$-band luminosity boosted up \citep{bekki01}, enough to fulfill our selection cut $M_B\leq -20$. And, (ii) the objects not picked up in the $B$-selected sample are earlier types dominated by an spheroidal component which, when subject to a major merger, do not distort sufficiently to be flagged as merger systems by our $A$-based criterion.  

\section{DISCUSSION}\label{discusion}
\subsection{The Importance of ML Method}\label{mlimpor}
Our merger fraction determination takes into account two potencial biases. First, we artificially redshift the images to take into account the pixelation and signal-to-noise degradation with redshift ($\S\S$~\ref{homoa}, \ref{avsz}). Second, we use a ML method to take into account the experimental errors in redshift and asymmetry (see LGB08). In Table~\ref{fftab} we compare merger fractions obtained with the same $B$-band catalog but with different methods. First, we explore the values obtained if we simply count galaxies without applying the ML method, i.e., if we mimic previous morphological merger determinations. In this case, tabulated uncertainties were estimated from Poisson statistics. We see that, if we use non-degraded asymmetries $A_0$ and the local merger criterion $A > 0.35$, we obtain a lower merger fraction than that from degraded asymmetries $A_{\rm f}$ and the $z=0.75$ merger criterion $A > 0.30$: 0.057 vs. 0.071 (20 vs. 25 merger systems, a $\sim 20$\% difference). Interestingly, when we apply the ML method over the two samples we obtain a similar merger fraction: 0.047 vs. 0.045, only a $4$\% difference. We comment on this fact later in this section.

\begin{deluxetable*}{lcc} 
\tabletypesize{}
\tablecolumns{5} 
\tablewidth{0pc}
\tablecaption{Merger fractions in the Groth Strip at $z = 0.6$\label{fftab}}
\tablehead{ 
Method & $f_{\rm m}^{\rm mph}(M_B\leq -20)$ & $f_{\rm m}^{\rm mph}(M_{K_{\rm s}} \leq -23.5)$
}
\startdata
$A_0 > 0.35$\tablenotemark{a}   				&   $0.057 \pm 0.013$  		&   $0.035 \pm 0.011$ \\
$A_0 > 0.35$\tablenotemark{a} + ML\tablenotemark{b}  		&   $0.047^{+0.014}_{-0.011}$   &   $0.032^{+0.013}_{-0.009}$\\
$A_{\rm f} > 0.30$\tablenotemark{c}   				&   $0.071 \pm 0.015$  		&   $0.045 \pm 0.012$ \\
${\bf A_{\rm f} > 0.30}$\tablenotemark{c} {\bf + ML}\tablenotemark{b} 	&   ${\bf 0.045^{+0.014}_{-0.011}}$ 	&   ${\bf 0.031^{+0.013}_{-0.009}}$\\
\enddata
\tablenotetext{a}{Using raw asymmetries $A_0$, and merger condition from C03 $(z
= 0)$.}
\tablenotetext{b}{Using ML method to determine the merger fraction.}
\tablenotetext{c}{Using $z=0.75$ asymmetries $A_{\rm f}$, and merger condition at $z = 0.75$ ($\S$~\ref{avsz}).}
\end{deluxetable*}

We also see that the effect of applying the ML method is more important when using $A_{\rm f}$ than when using raw asymmetries $A_0$: using ML with $A_{\rm f}$ leads to a merger fraction decrease from 0.071 to 0.045, a $\sim 60$\% difference, while, using $A_0$, it drops from 0.057 to 0.047, a $\sim 20$\% difference. This difference is due to the fact that errors in $A_{\rm f}$ are higher than errors in $A_0$: while $A_{\rm f}$ values are superior tracers of the asymmetries of the sample galaxies, they have higher errors $\sigma_{A}$ due to the uncertainties in the degradation process that transforms galaxy images from their original $z$ to $z_{\rm d} = 0.75$, and to the loss of signal-to-noise of those galaxies with $z < z_{\rm d}$.  
The mean $A_0$ error in $0.35 \leq z < 0.85$ sources is $\overline{\sigma_{A_0}} = 0.02$, while the mean $A_{\rm f}$ error in $0.35 \leq z < 0.85$ sources is $\overline{\sigma_{A_f}} = 0.05$. As noted in LGB08, the bigger the experimental errors, the more the classical histogram-based methods overestimate the merger fraction.  

For the $K_{\rm s}$-selected sample, the merger fractions depend on the usage of ML with similar trends to the $B$-selected sample.
In the $A_0$ case we change from 0.035 (11 merger systems) to 0.032, a $\sim 10$\% difference, while in the $A_{\rm f}$ case we change from 0.045 (14 merger systems) to 0.031, a $\sim 45$\% difference. As in the $B$-band case, the ML method values are similar: 0.032 vs. 0.031, a 3\% difference.

The comparison of merger fractions in Table~\ref{fftab} indicates that, in the redshift range of our study, histogram-based merger fraction determinations may be overestimated by up to $\sim$60\%, and that such overestimates are readily corrected by using ML method. Once ML methods are used, the impact of working with asymmetry values normalized to a given reference redshift is less than 4\%. Note that the effect of the experimental errors depends on the sample: our near infrared merger fraction is less affected by observational errors, but we do not know if this is a general trend, or a peculiarity of our samples.

In summary, if we do not use ML method to take into account the effect of observational errors, we overestimate the merger fraction by 10\%--60\%, in good agreement with the expected $\sim $10\%--30\% obtained in LGB08 with synthetic catalogs. On the other hand, as long as ML method is used, the effect of the pixelation and signal-to-noise degradation with redshift is less important, only a $\sim 4$\% effect. This fact, however, depends on the sample. A similar study in GOODS-S reveals that loss of information with redshift is also an important bias that needs to be treated to ensure accurate results (L\'opez-Sanjuan et al., in prep.).

\subsection{Comparison with Previous Morphological $B$-band Studies}\label{mphdisc}
In this section we compare our morphology-based merger fraction to other determinations in the literature. We restrict the comparison to works with $B$-band selected samples with well established luminosity limits, given the dependence of the merger fraction on the selection band, $\S$~\ref{bvsk}, and given that more luminous $B$-band samples tend to have higher merger fractions \citep{conselice03ff}. We exclude studies based on pair statistics, due to the progenitor bias (L08; \citealt{bell06}): each distorted galaxy in our sample is the final stage of the merger of two less luminous/massive galaxies. For example, assuming a 1:1 merger and neglecting star-formation, we need merge two $M_B = -19.25$ galaxies to obtain one high asymmetric $M_B = -20$ source. In addition, highly merger-triggered star-formation affects the $B$-band luminosity of the distorted source from $+0.5$ (dusty merger) to $-1.5$ magnitudes \citep[see][for details]{bekki01}.

In the following sections, we pay attention to how the information degradation and the experimental errors have been treated in previous works. The effect of the information degradation was addressed in all previous morphological studies at intermediate redshifts (e.g., \citealt{conselice03ff,conselice08,cassata05,kamp07}; L08), while the effect of experimental errors was studied in detail only by L08. Using synthetic catalogs in the same way as in this paper, $\S$~\ref{simu}, and in LGB08, they found similar trends: the experimental errors tend to overestimate the merger fraction. However, L08 did not apply any correction for this effect. Our work is the first in which this important bias is corrected. 

\subsubsection{The Merger Fraction at $z \sim 0.6$}
The most direct comparison is the work of \citet{conselice03ff}, who use the same asymmetry index as us. \citet{conselice03ff} lists a merger fraction $f_{\rm m}^{\rm mph}(z=0.6, M_B\leq -20) = 0.07$.  Their value is really an upper limit coming from 1 merger detection in 15 galaxies, i.e., $f_{\rm m}^{\rm mph} <  0.16$ assuming a Poisson distribution.

Other asymmetry works at intermediate redshift have been used different selection criteria than us: \citet{cassata05} obtain a merger fraction $f_{\rm m}^{\rm mph}(z=0.75) = 0.088^{+0.044}_{-0.026}$ in a $m_{K_{\rm s}} < 20$ selected sample. This value is a factor two higher than ours; we suspect that properly accounting for the experimental errors would reduce this discrepancy. \citet{bridge07} perform their asymmetry study on a 24$\mu$m-selected sample ($L_{IR} \geq 5.0 \times 10^{10}\ L_{\odot}$), finding $f_{\rm m}^{\rm mph}(z=0.75) \sim  0.16$.
Such high value are expected from the fact that strong star formation occurs in morphologically-distorted galaxies \citep{sanders88}.

Two other studies give the morphological merger fraction using methodologies related but not identical to ours: \citet{kamp07} measured the fraction of visually disturbed galaxies at $z \sim 0.7$, finding $f_{\rm m}^{\rm mph}(z=0.7, M_B\leq -19.15) = 0.024$, while L08 used the morphological indices $G$ and $M_{20}$ to obtain $f_{\rm m}^{\rm mph}(z=0.7, M_B\leq -19.75) = 0.07^{+0.06}_{-0.01}$. In next section we use L08 data to constraint the merger fraction evolution, showing the difficulties entrained when combining merger fractions from different methodologies.

In summary, previous morphological works in $B$-band selected samples suggest that $f_{\rm m}^{\rm mph}(z \sim 0.7) \sim 0.07$, higher than our value of 0.045. We recover their value ($f_{\rm m}^{\rm mph} =  0.071$) when mimicking their methodology, i.e., when we apply the merger condition $A_{\rm f} > 0.30$ without using ML method. This suggests that merger fraction values around $f_{\rm m}^{\rm mph}(z \sim 0.7) \sim 0.07$ may be overestimated by $\sim$50\% due to not properly accounting for sources spilling over to neighboring bins.

\subsection{Morphological Merger Fraction Evolution}\label{ffevol}
We now constrain the evolution of the merger fraction with redshift by combining our result at $z=0.6$ with those of L08 at $z=0.9$, and \citet{depropris07} at $z=0.07$.

To combine our results with L08 we check, first, that their luminosity-dependent sample selection, $M_B\leq -18.83 - 1.3z$\footnote{This luminosity cut selects $L_{B} > 0.4L_{B}^{*}$ galaxies and takes into account the evolution of $L_{B}^{*}$ with redshift \citep{faber07}.}, matches our luminosity selection at $z \sim 0.9$.  We then note that L08 use different morphological indices than ours, namely $G$ and $M_{20}$. This introduces a time factor $\Delta T_{\rm m}$ that relates the time that distorted sources fulfill the merger criteria in each of the methods: $\Delta T_{\rm m} = T_{{\rm m},A}/T_{{\rm m},GM_{20}}$. The different merger time-scales can be constrained with merger simulations. \citet{lotz08t} performed N-body/hydro-dynamical simulations of equal-mass gas-rich disk mergers and studied the time-scales $T_{{\rm m},A}$ and $T_{{\rm m},GM_{20}}$ as a function of the merger parameters.  
Their results suggest that $\Delta T_{\rm m} = 1.5 \pm 0.5$. In addition, and taking into account the results of $\S$~\ref{mlimpor}, we estimate a 20\% overestimate on the merger fraction due to morphological index errors ($\langle \sigma_G \rangle \sim 0.02$). With all these considerations, the wet merger fraction at $z = 0.9$ from L08, $f_{\rm m}^{\rm mph}(z=0.9) = 0.06 \pm 0.01$\footnote{L08 find 41/685 blue (wet) mergers and 15/685 red (dry) mergers at z = 0.9.} becomes 
\begin{equation}
f_{\rm m}^{\rm mph,L08}(z=0.9, M_B\leq -20) = 0.075^{+0.03}_{-0.03}
\end{equation}
in our methodology. The error is dominated by the uncertainty in the time-scale factor $\Delta T_{\rm m}$.

\citet{depropris07} provide a suitable low-redshift determination of the merger fraction using asymmetries, but their result cannot be directly combined with ours since they extend to fainter absolute magnitudes, namely $M_B\lesssim -19$. \citet{conselice03ff} results suggest that $f_{\rm m}^{\rm mph}(M_B\leq -20) \sim 1.5 f_{\rm m}^{\rm mph}(M_B\leq -19)$ by asymmetries, while pair statistics at that redshift also suggest that $f_{\rm m}^{\rm mph}(M_B\leq -20) \gtrsim f_{\rm m}^{\rm mph}(M_B\leq -19)$ \citep[][]{patton08}\footnote{\citet{patton08} perform their study in $M_r$ (SDSS) absolute magnitude samples. We take $B$-$r$ = 1.25 \citep{fukugita95} to obtain the equivalent $M_B$ samples.}. Because of this, we apply the C03 factor to \citet{depropris07} results, and estimate that 
\begin{equation}
f_{\rm m}^{\rm mph,P07}(z = 0.07, M_B\leq -20) = 0.014^{+0.003}_{-0.003}.
\end{equation}

\begin{figure}[t]
\plotone{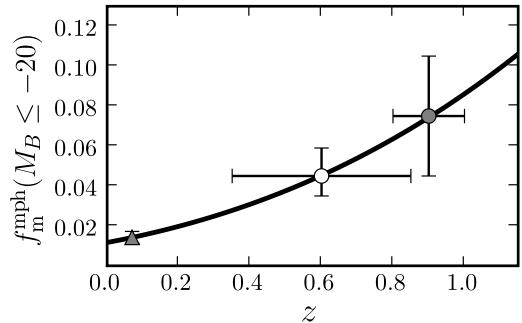}
\caption{Morphological merger fraction in function of redshift for $M_B \leq -20$ galaxies. The data are \citet[][{\it gray triangle}]{depropris07}, L08 ({\it gray dot}), and this work ({\it white dot}). The black solid line is the best fit to data, $f_{\rm m}^{\rm mph}(z,M_B\leq -20)= 0.012(1+z)^{2.9}$.}
\label{ffm20}
\end{figure}

We parameterize the merger fraction as $f_{\rm m}^{\rm mph}(z) $ = $ f_{\rm m}^{\rm mph}(0)$ $(1+z)^m$ \cite[e.g.,][]{lefevre00}, and perform a weighted least-squares fit to the data, obtaining that 
\begin{equation}
f_{\rm m}^{\rm mph}(z,M_B\leq -20) = (0.012 \pm 0.004)(1+z)^{2.9 \pm 0.8}.\label{fmfit}
\end{equation}
Figure~\ref{ffm20} shows the three merger fractions from \citet[][{\it gray triangle}]{depropris07}, L08 ({\it filled circle}) and this work ({\it open circle}). The black solid line is the weighted least-squares fit to the data, equation~(\ref{fmfit}). This result supports the idea of important evolution of the merger fraction ($m \gtrsim 2$) from $z \sim 1$ to the present \citep[e.g.,][]{lefevre00,lavery04,cassata05,bell06,kamp07,kar07,rawat08,hsieh08}.
This evolution, however, do not imply that disk-disk major mergers were important in galaxy evolution since $z \sim 1$: the local merger fraction is low, $\sim0.01$, and,  despite of the $m = 2.9$ evolution, the merger fractions remain below $0.1$ up to $z \sim 1$. It is only for $z > 1$ that our high exponent translates into a high merger fraction: $f_{\rm m}^{\rm mph} \sim 0.2$ at $z \sim 1.5$, extrapolating our fit.  At those redshifts, major disk-disk mergers are sufficiently frequent that they may be important for galaxy evolution \citep{conselice06,conselice08}. We study the importance of this type of mergers since $z = 1$ in the next section.

It is worth recalling that our $m = 2.9$ do not necessary imply an important decrease with cosmic time of the \textit{total} merger fraction since $z=1$.
Our merger criterion is only sensitive to disk-disk major mergers. The pair major merger studies of \citet{lin08}; and \citet{deravel08} show that the wet merger fraction evolution is higher than the total (dry+wet) merger fraction evolution. Our study cannot rule out an increasing importance of dry mergers at $z<1$.  

\subsubsection{Major merger remnants since $z \sim 1$}\label{frem}
Following \citet{patton00}, the fraction of present-day $M_B\leq -20$ galaxies that have undergone a disk-disk major merger since a given redshift is:
\begin{equation}
f_{\rm rem} = 1 - \prod^{N}_{j=1}\frac{1 - f_{\rm m}^{\rm mph}(z_j)}{1 - 0.5f_{\rm m}^{\rm mph}(z_j)},
\end{equation}
where $f_{\rm m}^{\rm mph}(z)$ is the merger fraction at redshift $z$, parameterized as $ f_{\rm m}^{\rm mph}(0)$ $(1+z)^m$, $z_j$ corresponds to a look-back time of $t = jT_{{\rm m},A}$, and $N$ is the number of $t$ steps since a given redshift z.
We use the merger fraction parameters obtained in the previous section, and take two values for the merger time-scale: $T_{{\rm m},A} = 0.6$ Gyr \citep[][from N-body/hydro-dynamical equal-mass merger simulations]{lotz08t}, and $T_{{\rm m},A} = 0.35$ Gyr \citep[][from N-body major merger simulations]{conselice06}.

We find that the disk-disk major merger remnant fraction since $z = 1$ is $f_{\rm rem} \sim 20$\% for $T_{{\rm m},A} = 0.6$ Gyr, and $f_{\rm rem} \sim 35$\% for $T_{{\rm m},A} = 0.35$ Gyr. 

\subsection{The Major Merger Rate at $z=0.6$}
We now use the derive merger fraction for the $K_{\rm s}$ sample to study the merger rate and its variation with galaxy mass.
We define the major merger rate $\Re_{\rm m} (z,M_{K_{\rm s}})$ as:
\begin{equation}
\Re_{\rm m} (z,M_{K_{\rm s}}) = n(z,M_{K_{\rm s}}) f_{\rm m}^{\rm mph}(z,M_{K_{\rm s}}) T_{{\rm m},A}^{-1},
\end{equation}
where $n(z,M_{K_{\rm s}})$ is the comoving number density of galaxies at redshift $z$ brighter that $M_{K_{\rm s}}$, and $T_{{\rm m}, A}$ is the merger time-scale, which we take as $T_{{\rm m}, A} = 0.35 - 0.6$ Gyr ($\S$~\ref{frem}). To obtain $n(z,M_{K_{\rm s}})$ we use the \citet{cirasuolo08} luminosity function, that yields $n(0.6,-23.5) = 0.0025\ {\rm Mpc}^{-3}$. With these values, we obtain 
\begin{equation}
\Re_{\rm m} (0.6, -23.5) = 1.6^{+0.9}_{-0.6} \times 10^{-4}\ {\rm Mpc}^{-3}\ {\rm Gyr}^{-1}. 
\end{equation}
The error takes into account both merger fraction and merger time-scale uncertainties.

\begin{figure}[t]
\plotone{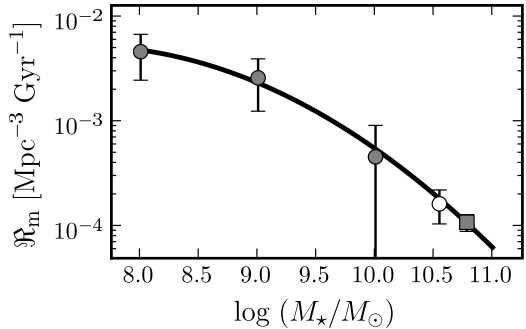}
\caption{Major merger rate in function of stellar mass at $z = 0.6$. The data are from \citet[][{\it gray dots}]{conselice08}, \citet[][{\it gray square}, error bars are smaller than the symbol size]{bell06}, and this work ({\it white dot}). The solid line is the best equation~(\ref{mexp}) fit to the data.}
\label{revsm}
\end{figure}

Adopting a constant mass-to-light ratio in the $M_{K_{\rm s}}$ band of $M_{\star}/L_{K} = 0.7$ \citep{drory04,arnouts07}, our $M_{K_{\rm s}} \leq -23.5$ luminosity selection corresponds to a $M_{\star} \gtrsim 3.5 \times 10^{10}\ M_{\odot}$ mass selection. Our inferred merger rate is shown against galaxy stellar mass in Figure~\ref{revsm}, and listed in Table~\ref{retab}, together with literature values for the major merger rate at $z=0.6$ at various limiting masses. Errors take into account both merger fraction and merger time-scale uncertainties. We can see that the merger rate at $z=0.6$ decreases with mass. We find that the variation with $M_{\star}$ is well described by
\begin{equation}\label{mexp}
\Re_{\rm m} (M) = \Re_0 {\rm e}^{\beta M^2},
\end{equation}
where $M = \log(M_{\star}/M_{0})$, and $M_{0}$, $\Re_0$, and $\beta$ are parameters to fit. The best $\chi^2$ fit to the data yields $M_{0} = 3.2 \times 10^{7}\ M_{\odot}$, $\Re_0 = 5.3 \times 10^{-3}\ {\rm Mpc}^{-3}\ {\rm Gyr}^{-1}$, and $\beta = -0.36$. Note that \citet{bell06} merger criterion is sensitive to both disk and spheroid mergers, while asymmetry studies are only to disk-disk mergers \citep{conselice06}. This implies that the data from \citet{bell06} must be higher than the disk-disk merger rate. If we repeat our analysis without the $6 \times 10^{7}\ M_{\odot}$ point, we obtain $M_{0} = 6.3 \times 10^{10}\ M_{\odot}$, $\Re_0 = 4.8 \times 10^{-3}\ {\rm Mpc}^{-3}\ {\rm Gyr}^{-1}$, and $\beta = -0.45$. With these values, the inferred disk-disk merger rate at $6 \times 10^{10}\ M_{\odot}$ is 80\% of the \citet{bell06} value.

In addition, our infered major merger fraction for $M_{\star} \gtrsim 3.5 \times 10^{10}\ M_{\odot}$ galaxies, $f^{\rm mph}_{\rm m} = 0.031^{+0.013}_{-0.009}$, is in agreement with the visual estimate of the morphological major merger fraction of $M_{\star} \geq 2.5 \times 10^{10}\ M_{\odot}$ galaxies in the range $z \in [0.34, 0.8]$, $f^{\rm mph}_{\rm m} \sim 0.02 \pm 0.01$ \citep{jogee08}.

\begin{deluxetable}{lcc}
\tabletypesize{}
\tablecolumns{5} 
\tablewidth{0pc} 
\tablecaption{Merger Rates at $z = 0.6$\label{retab}}
\tablehead{ 
Reference & Stellar mass & $\Re_{\rm m}(z,M_{\star})$\\
& ($M_{\odot})$ & $(10^{-4}\ {\rm Mpc}^{-3}\ {\rm Gyr}^{-1})$}
\startdata
\citet{conselice08}   &  $10^{8}$              &   $46.9 \pm 23.7$   \\
\citet{conselice08}   &  $10^{9}$              &   $26.4 \pm 14.7$   \\
\citet{conselice08}   &  $10^{10}$             &   $4.6  \pm 4.6$    \\
This work             &  $3.5 \times 10^{10}$  &   $1.6^{+0.9}_{-0.6}$    \\
\citet{bell06}	      &  $6 \times 10^{10}$    &   $1.1  \pm 0.2$
\enddata 
\end{deluxetable}

\section{CONCLUSIONS}\label{conclusion}
From Groth strip survey images, we provide a robust determination of the disk-disk major merger fraction based on morphological criteria. We have quantified and corrected for the bias due to varying spatial resolution and image depth with redshift, by artificially redshifting the galaxy images to a common reference redshift of $z_{\rm d} = 0.75$.  More importantly, we successfully accounted for spill-over of sources to neighboring bins caused by the errors in asymmetry indices and in $z_{\rm phot}$, through the use of a ML method developed in LGB08.  The merger fractions for the $B$-selected and $K_{\rm s}$-selected samples are, respectively, 
$$
f_{\rm m}^{\rm mph}(z = 0.6, M_B\leq -20) = 0.045^{+0.014}_{-0.011},
$$
$$
f_{\rm m}^{\rm mph}(z = 0.6, M_{K_{\rm s}} \leq -23.5) = 0.031^{+0.013}_{-0.009}.
$$

The effect of the experimental errors is the dominant observational bias in our study: without the ML method we overestimate the galaxy merger fraction by up to 60\%. In comparison, the loss of information with redshift only biases the results by $\sim 4$\%, as long as ML is used to account for the experimental errors.

Parameterizing the merger fraction as $f_{\rm m}^{\rm mph} = f_{\rm m}^{\rm mph}(0) (1+z)^m$, we obtain $m = 2.9 \pm 0.8$, $f_{\rm m}^{\rm mph}(0) = 0.012 \pm 0.004$. With these values, we infer that only $20-35$\% of present-day $M_B \leq -20$ galaxies have undergone a disk-disk major merger since $z \sim 1$.

We use the $M_{K_{\rm s}}$-band merger fraction to obtain the major merger rate at $z = 0.6$. 
 Assuming a constant mass-to-light ratio, we obtain $\Re_{\rm m}(0.6, 3.5 \times 10^{10} M_{\odot}) = 1.6^{+0.9}_{-0.6} \times 10^{-4}\ {\rm Mpc}^{-3}\ {\rm Gyr}^{-1}$. We compare our results with previous merger rates at that redshift, showing that the merger rate rapidly decreases with mass, such that the rate at $M_{\star} = 10^{10.5}\ M_{\odot}$ is 10 times lower than that at $M_{\star} = 10^{9}\ M_{\odot}$.

\acknowledgements
We dedicate this paper to the memory of our six IAC colleagues and friends who met with a fatal accident in Piedra de los Cochinos, Tenerife, in February 2007, with a special thanks to Maurizio Panniello, whose teachings of \texttt{python} were so important for this paper.
This work was supported by the Spanish Programa Nacional de Astronom\'{\i}a y Astrof\'{\i}sica through project number AYA2006-12955.

{\it Facilities:} \facility{HST (WFPC2)}; \facility{ING:Herschel (INGRID)}; \facility{ING:Newton (WFC)}

\end{document}